\documentclass[10pt,letterpaper,twocolumn]{article}

\usepackage[T1]{fontenc}
\usepackage[utf8]{inputenc}
\usepackage{lmodern}
\usepackage[margin=0.75in,columnsep=0.25in]{geometry}
\usepackage{amsmath,amssymb}
\usepackage{array,booktabs,graphicx,placeins}
\usepackage[numbers,sort&compress]{natbib}
\usepackage{microtype}
\usepackage{xurl}
\usepackage[hidelinks]{hyperref}
\usepackage[font=small,labelfont=bf]{caption}

\providecommand{\Description}[1]{}

\graphicspath{{figures/}}
\hypersetup{
  pdftitle={SignalAudit-QD: A Reproducible Audit Resource for External Signals in AI-Query Forecasting},
  pdfauthor={Amir Rafe, Iman Jamshidi, Gourab Sil, Subasish Das},
  pdfsubject={Temporal query forecasting and external-signal evaluation},
  pdfkeywords={query forecasting, Google Trends, external signals, reproducibility}
}

\title{SignalAudit-QD: A Reproducible Audit Resource for External Signals in AI-Query Forecasting}
\author{%
  \normalsize
  \href{https://orcid.org/0000-0002-4089-2088}{Amir Rafe}\textsuperscript{1}\quad
  \href{https://orcid.org/0009-0000-5949-8833}{Iman Jamshidi}\textsuperscript{1}\quad
  \href{https://orcid.org/0000-0002-6930-4606}{Gourab Sil}\textsuperscript{2}\quad
  \href{https://orcid.org/0000-0002-1671-2753}{Subasish Das}\textsuperscript{1}\\[0.4em]
  \small\textsuperscript{1}Texas State University, San Marcos, USA\\
  \small\textsuperscript{2}Indian Institute of Technology Indore, Indore, India\\[0.3em]
  \small\texttt{amir.rafe@txstate.edu}, \texttt{ijamshidi@txstate.edu}\\
  \small\texttt{gourabsil@iiti.ac.in}, \texttt{subasish@txstate.edu}
}
\date{}

\begin{document}
\twocolumn[{
\begin{@twocolumnfalse}
\maketitle
\begin{center}
\begin{minipage}{0.95\textwidth}
\begin{abstract}
SignalAudit-QD is a reproducible audit resource for testing whether a dated public signal adds information beyond query history in a fixed 20-term U.S. AI panel. It aligns two archived Google Trends extractions with eight USPTO AI-patent publication streams and releases origin-indexed predictions, paired losses, provenance, executable tests, and tamper-evident run manifests. Both inputs are retrospective reconstructions obtained after the modeled 2020--2023 origins; the task is therefore a pseudo-origin assay within frozen measurement vintages, not a historical deployment simulation. In the designated one-month comparison, adding a two-component patent block to a level ridge changes macro mean absolute error by $+0.0012$ Trends points, with conditional 95\% moving-block interval $[-0.0363,+0.0387]$. This establishes neither aggregate benefit nor equivalence. Alternative dimensions, semantic mapping, patent definitions, a second extraction, and a three-month horizon remain unresolved; persistence-anchored and nonlinear hosts instead show model-specific worsening, tiny for the anchored fit. Matched timing controls expose availability assumptions, while a planted-signal assay detects only the largest tested injection. A within-query high-interest alert demonstrates the released interface but not online utility. The contribution is a bounded negative case study and an extensible evidence schema for auditing external-signal claims in temporal information retrieval.
\end{abstract}

\end{minipage}
\end{center}
\vspace{1em}
\end{@twocolumnfalse}
}]

\section{Introduction}
\label{sec:introduction}

\begin{figure*}[t]
    \centering
    \includegraphics[width=0.9\textwidth]{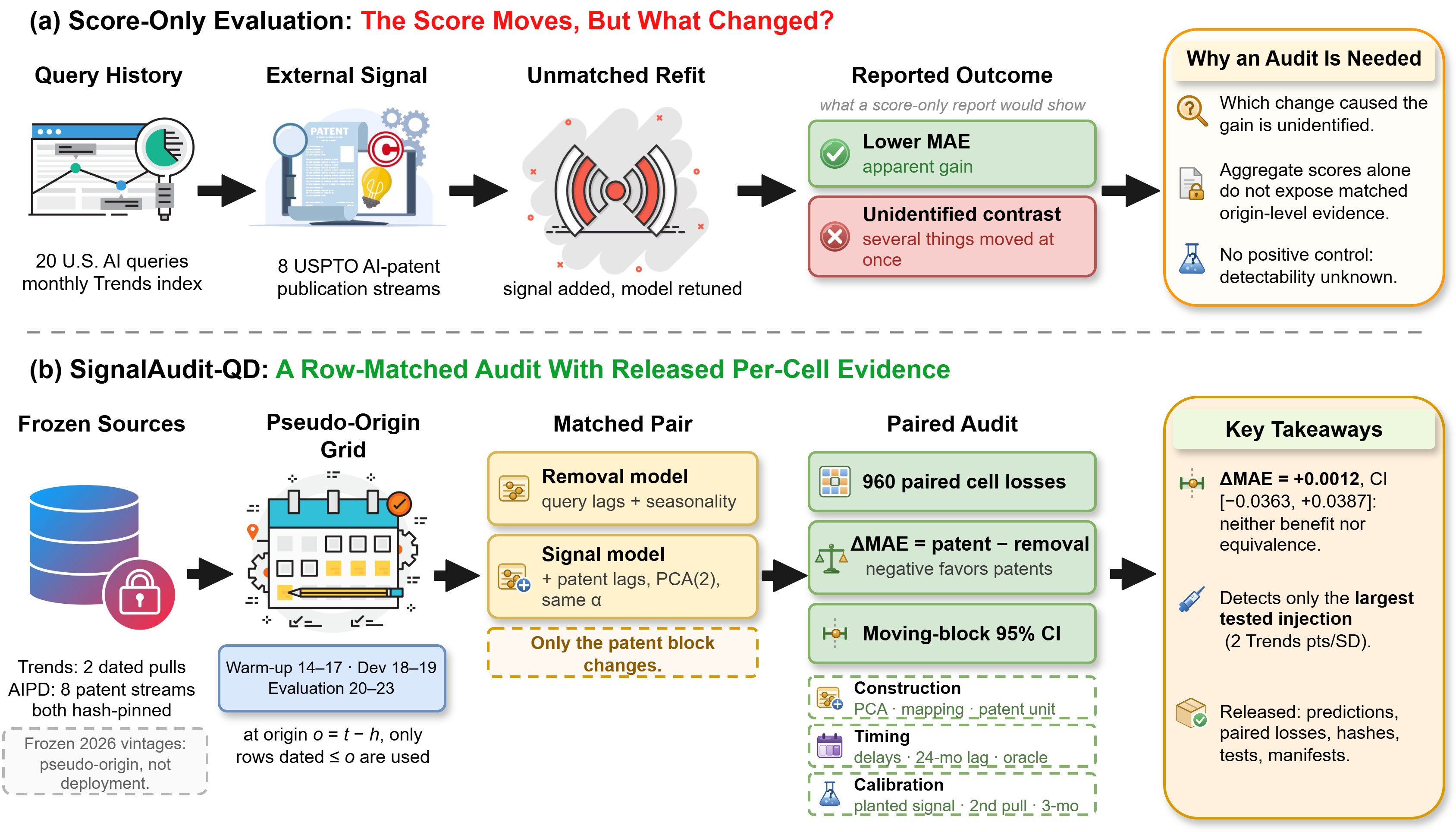}
    \caption{Overview of SignalAudit-QD. (a) Score-only evaluation cannot isolate whether a performance change is caused by the external signal or other modeling changes. (b) SignalAudit-QD uses frozen sources and a fixed pseudo-origin grid to construct row-matched signal and removal models that differ only in the patent block, producing paired losses for audit and calibration.}
    \Description{Two-panel overview contrasting a score-only evaluation with the SignalAudit-QD row-matched auditing pipeline. The proposed pipeline uses frozen query and patent sources, pseudo-origin evaluation, matched signal and removal models, paired cell-level losses, and construction, timing, and calibration checks.}
    \label{fig:overview}
\end{figure*}

Forecasts of query interest can support time-sensitive query autocompletion, monitor emerging demand, and inform decisions about when query-specific ranking features should be refreshed \citep{shokouhi2012timesensitive}. External signals may improve such forecasts, but their incremental value is easy to overstate: adding a covariate may also alter the usable history, preprocessing transformations, regularization, or target vintage. Consequently, a difference in aggregate forecasting error does not reveal whether the external signal contributed new information or whether other elements of the forecasting experiment changed. Reproducible evaluation in temporal information retrieval therefore requires a matched and inspectable comparison that answers a basic question: \emph{what changed between the forecasts?} SignalAudit-QD operationalizes this requirement by comparing models fitted with and without the external signal on exactly matched query--origin rows. As illustrated in Figure~\ref{fig:overview}, conventional score-only evaluation leaves the source of an apparent performance difference unresolved, whereas SignalAudit-QD releases query--origin-level predictions, paired losses, provenance records, executable tests, and run manifests that make each comparison directly inspectable.

Google Trends offers public aggregate trajectories without user records, but the target is fragile. It is a sampled, normalized 0--100 index rather than a query count; terms requested independently support within-term temporal comparison, not absolute volume comparison across terms \citep{googletrendsfaq}. Repeating a historical request can also change values \citep{cebrian2024trends}. A forecast benchmark built on Trends must freeze each extraction, document low-volume zeros, and distinguish event time from the later date on which a retrospective series was retrieved.

We instantiate this problem with a deliberately narrow question: does a public block of U.S. AI-patent publication signals add forecast information beyond the same AI-query-history learner? The USPTO Artificial Intelligence Patent Dataset (AIPD) supplies eight monthly AI-component streams through 2023 \citep{pairolero2025aipd}. The proposed information pathway is weak but testable: patent publication is a dated part of the public AI information supply that may precede broader attention. It is not invention time, adoption, exposure, or a causal treatment.

\begin{figure*}[t]
    \centering
    \includegraphics[width=0.9\textwidth]{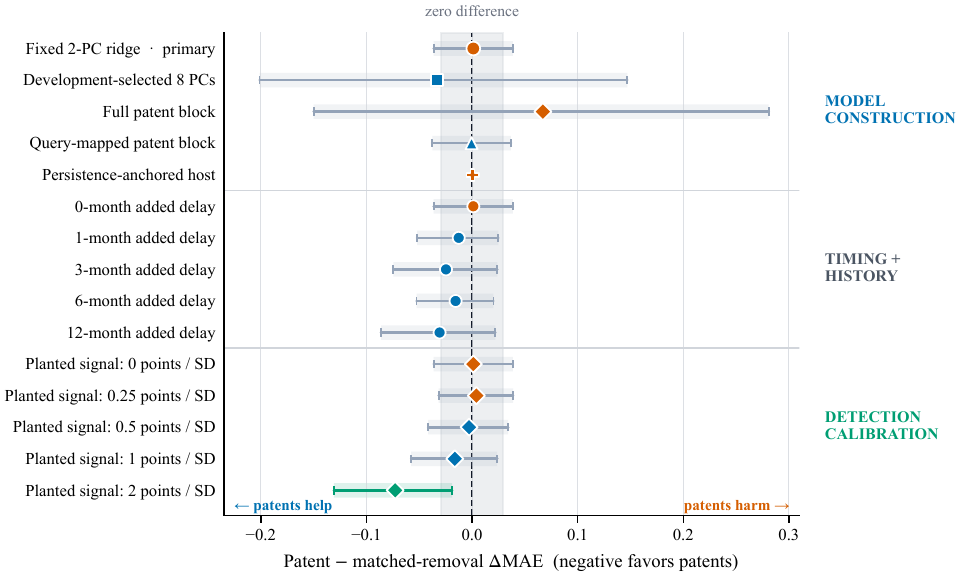}
    \caption{Evidence atlas for the canonical one-month task. Each row is a designated patent-minus-matched-removal $\Delta$MAE estimate with a conditional 95\% three-month moving-block interval; negative values favor patents. Model-construction and timing--history checks remain unresolved because their intervals cross zero. The 0-month timing row restates the designated primary contrast to anchor the timing sequence. In the planted-signal calibration, only the 2-point-per-standard-deviation injection is detected below zero. The gray corridor is $\pm1\%$ of persistence MAE, a descriptive practical reference rather than an equivalence threshold.}
    \Description{An evidence-atlas forest plot with 15 rows on a shared horizontal effect scale. The top five rows are matched model constructions, the middle five are zero- to twelve-month timing and history pathways, and the bottom five are planted-signal calibration levels. Each point has a conditional interval. A central gray corridor marks a descriptive practical reference, and only the strongest planted signal has an interval entirely below zero.}
    \label{fig:signal-audit}
\end{figure*}

SignalAudit-QD is the resulting case-study resource and evidence package. Its designated primary pair changes only the patent block whose publication dates are no later than the pseudo-origin; the 2026 source vintages themselves are held fixed and are not claimed to have existed historically. The release also contains row-matched latency variants, a non-wrapping 24-month lag stress test, a future-information oracle, development-selected PCA width, a transparent query-to-patent map, a persistence-anchored host, alternative patent aggregations, and a planted-signal assay. Figure~\ref{fig:signal-audit} organizes the designated construction, timing, and calibration checks on one evidence scale. The repaired lag control no longer yields the earlier apparent harmful effect; this is precisely why executable audit transformations belong in the contribution.


Our contributions are:

\begin{enumerate}
    \item \textbf{A versioned temporal-IR resource.} Two dated full-history Trends pulls for a fixed 20-query U.S. AI panel are aligned with eight AIPD streams, an exact query catalog, source hashes, per-file terms, and tamper-evident run manifests.
    \item \textbf{A matched signal-assay protocol.} Pseudo-origin predictions and paired cell/month losses are released for level and persistence-anchored hosts, with matched-row latency, PCA-width, semantic mapping, patent-definition, oracle, lag, and positive-control diagnostics.
    \item \textbf{A bounded empirical case study and IR proxy task.} The primary contrast is unresolved around zero; the same bounded conclusion holds across two archived target extractions and at three months. A within-query high-interest alert exposes a possible refresh interface at one fixed operating point without claiming downstream utility.
\end{enumerate}

\section{Prior Research and Study Contribution}
\label{sec:related}

Forecasting query demand is not new. Earlier work characterized temporal query classes, unified web-query prediction, and exploited temporal patterns for autocompletion \citep{liu2008query,kulkarni2011temporal,jiang2017temporal}. Time-sensitive query autocompletion used predicted frequency to change an IR ranking decision, but evaluated commercial logs that were not released as a public benchmark \citep{shokouhi2012timesensitive}. These task-nearest studies do not release dated target vintages plus origin-indexed signal/removal evidence. GoogleTrendArchive instead prospectively captures 7.6 million real-time Trending Now episodes across 125 countries; it supports inductive study of surge events rather than longitudinal trajectories for a fixed term panel \citep{urman2026googletrendarchive}. Google Cloud's BigQuery tables likewise expose rolling top and rising terms \citep{googlebigquery2021}.

Measurement resources solve complementary problems. G-TAB calibrates terms through an anchor bank to mitigate 0--100 rounding and cross-request scale incompatibility \citep{west2020calibration}. Google's application-gated Trends API alpha provides consistently scaled observations over a rolling five-year window, but is not yet a frozen evaluation resource \citep{googletrendsapi2026}. Repeated public-interface pulls can vary because of sampling and normalization, and recent statistical preprocessing work treats censoring, noise, and trend explicitly \citep{cebrian2024trends,djorno2026restoring}. SignalAudit-QD preserves dated pulls and exposes their disagreement; it does not solve calibration or recover latent search counts.

Exogenous-variable forecasting is also mature. TimeXer evaluates an exogenous-input architecture on twelve forecasting benchmarks \citep{wang2024timexer}; the Monash archive and TFB provide much broader datasets and evaluation interfaces \citep{godahewa2021monash,qiu2024tfb}. In IR, QPP4CS predicts retrieval effectiveness rather than future query interest \citep{meng2023qpp}. Table~\ref{tab:closest-resources} therefore positions SignalAudit-QD as a compact \emph{retrospective signal-assay resource}, not a new forecasting architecture, general search archive, calibration method, or query-performance predictor.

\begin{table*}[t]
    \caption{Closest-resource comparison. ``Origin evidence'' means released rolling-origin predictions or paired losses.}
    \label{tab:closest-resources}
    \centering
    \footnotesize
    \setlength{\tabcolsep}{3.4pt}
    \begin{tabular}{@{}p{0.18\textwidth}p{0.27\textwidth}p{0.14\textwidth}p{0.15\textwidth}p{0.19\textwidth}@{}}
        \toprule
        Resource & Task and scale & Target-vintage evidence & Origin evidence & Complementary boundary \\
        \midrule
        Shokouhi--Radinsky \citep{shokouhi2012timesensitive} & Autocomplete frequency; 846K private log queries & No & Not released & Operational ranking use; no public demand panel \\
        GoogleTrendArchive \citep{urman2026googletrendarchive} & 7.6M Trending Now episodes; 125 countries & Prospective 2024--2026 collection & No forecasts & Inductive surge archive; not fixed-term trajectories \\
        G-TAB \citep{west2020calibration} & Cross-query Trends calibration via anchor bank & Calibration reference & No forecasts & Repairs rounding/scaling; not a forecast audit \\
        Trends API alpha \citep{googletrendsapi2026} & Consistently scaled, rolling five-year search interest & Current rolling service & No forecasts & Gated source infrastructure; not a frozen benchmark \\
        TimeXer \citep{wang2024timexer} & Exogenous forecasting; 12 benchmarks & No source-vintage layer & Experiment outputs & Architecture ablations; no matched availability assay \\
        QPP4CS \citep{meng2023qpp} & Retrieval-quality prediction; three collections & Not applicable & Per-query predictions & Different target; no temporal signal audit \\
        Monash / TFB \citep{godahewa2021monash,qiu2024tfb} & General archives; 8,068 univariate plus multivariate sets in TFB & Dataset versions, not repeat pulls & Baselines/adapters & Broad model comparison; no source-specific assay \\
        \textbf{SignalAudit-QD} & 20 fixed U.S. AI strings; two horizons; one patent family & \textbf{Two dated full-history pulls} & \textbf{Predictions and paired cell/month losses} & \textbf{Matched removal, timing, mapping, construct, and sensitivity controls} \\
        \bottomrule
    \end{tabular}
\end{table*}

Among the task-nearest studies and public resources surveyed here, we found no other resource combining a fixed query-interest panel across multiple dated full-history pulls with released origin-indexed signal/removal losses and explicit timing and planted-signal controls. This scoped, conjunctive distinction is the contribution, not an exhaustive priority claim for query forecasting, Trends data, exogenous modeling, or time-series benchmarking.

\section{SignalAudit-QD Resource}
\label{sec:resource}
\subsection{Query targets, sampling frame, and zeros}

The panel comprises 20 U.S. Google search strings requested independently over January 2014--December 2025. The 120-month 2014--2023 intersection with patent coverage is modeled. The strings were inherited from the predecessor data collection and fixed in the canonical 2026-04-07 extraction before the SignalAudit-QD protocol was designed. The original project did not archive a candidate universe, annotator record, or deterministic inclusion rule; we therefore treat this as a fixed purposive case-study panel, not a reproducible or representative sample of information needs. The release makes that limitation auditable. \texttt{query\_catalog.csv} records request order, exact spelling, first nonzero month, pre-evaluation and evaluation zero shares, and a post-hoc descriptive category that was not used for selection; each vintage's metadata records geography, language, timezone, property, window, client, and response processing. Eight terms have at least 50\% zeros before evaluation. Trends documents that a zero can reflect insufficient search volume, not observed zero demand \citep{googletrendsfaq}; our low-base sensitivity excludes these eight terms rather than imputing unsupported latent counts.

Every term was collected in its own request. We use each 0--100 index only through time within that term. Two complete processed monthly panels, dated 2026-04-07 and 2026-08-25, are retained with wide/long files and metadata; raw HTTP responses and April client version were not preserved. The panels measure robustness across two pulls, not separate sampling, renormalization, and revision effects.

\begin{figure*}[t]
    \centering
    \includegraphics[width=0.98\textwidth]{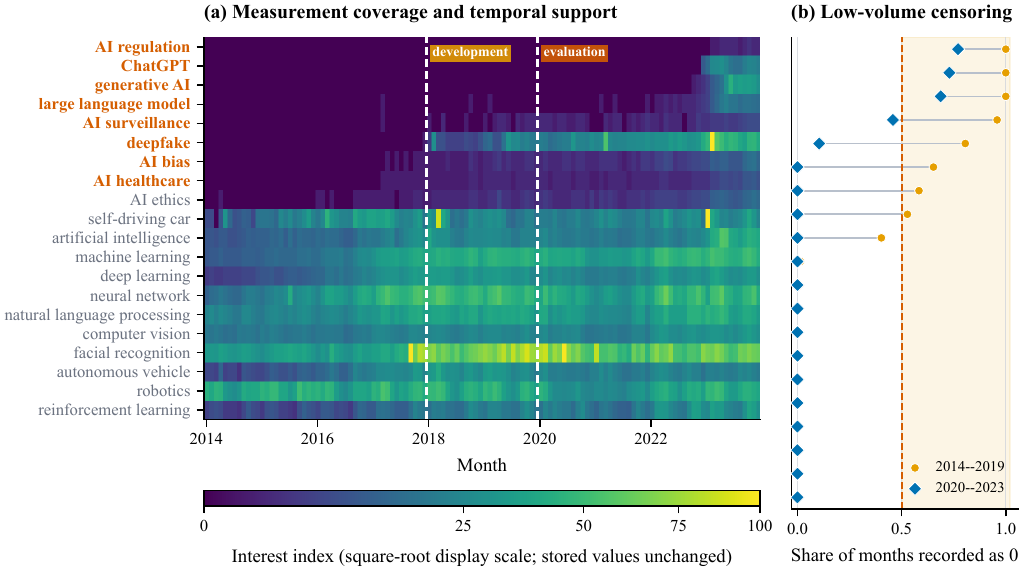}
    \caption{Measurement support in the fixed query panel. (a) The complete 2014--2023 canonical target matrix, ordered by pre-evaluation zero share; a square-root display transform reveals low values but does not alter stored data or analysis. Dashed lines mark development and evaluation boundaries. (b) Recorded-zero shares before and during evaluation. Bold orange labels satisfy the released low-base exclusion rule (pre-evaluation share $\geq0.5$). Google Trends zeros may indicate insufficient volume, so this is a censoring diagnostic rather than a count of absent searches.}
    \Description{A two-panel figure shows a 20-query by 120-month heatmap and a paired dot plot of zero shares. Eight emerging or low-base terms are highlighted; several are nearly all zero before evaluation and become measurable only late in the series.}
    \label{fig:panel-measurement}
\end{figure*}

Figure~\ref{fig:panel-measurement} makes the measurement boundary visible instead of hiding it in a summary statistic. The eight excluded terms are not simply low-amplitude versions of mature queries: several have no recorded pre-evaluation support and remain heavily zero-coded during evaluation. Conversely, long-running technical terms supply dense temporal histories but still inhabit separately normalized scales. This pattern motivates the released per-term diagnostics, the within-query alert definition, and the rule against pooling levels as comparable demand volumes.

\begin{table*}[t]
    \caption{SignalAudit-QD resource card. MIT covers author software only; \texttt{LICENSE\_MATRIX.csv} records terms and notices for every file class.}
    \label{tab:resource-card}
    \centering
    \small
    \setlength{\tabcolsep}{4pt}
    \begin{tabular}{@{}p{0.16\textwidth}p{0.14\textwidth}p{0.17\textwidth}p{0.20\textwidth}p{0.19\textwidth}@{}}
        \toprule
        Component & Dimensions & Coverage & Unit / interpretation & Released evidence \\
        \midrule
        Trends pulls & $2\times144\times20$ & 2014-01--2025-12; U.S.; monthly & Independently normalized sampled interest & Wide/long CSV, dated metadata, hashes, collector \\
        Modeling slice & $2\times120\times20$ & 2014-01--2023-12 & Retrospective warm-up, development, evaluation & Exact split, query catalog, zero audit \\
        AIPD panels & $120\times8$ each & 2014-01--2023-12; publication month & Document-score sum; grants, pre-grants, and mean/document variants & Builder, source hash, row/type counts, metadata \\
        Forecast archive & 105,280 rows & Two pulls, two horizons, 28 model variants & Actual, forecast, loss, origin, training rows, timing/vintage flags & Cell/month pairs, metrics, intervals, selected settings \\
        Verification & Executable suite & Timing, matching, PCA fit, mapping, outputs, secondary audits & Leakage, identity, malformed-adapter, and golden-result checks & Direct pins, realized tool logs, run/release ledgers \\
        \bottomrule
    \end{tabular}
\end{table*}

\subsection{Patent constructions and semantic map}

The primary external panel is reconstructed from AIPD 2023 revision \texttt{2023\_r1} \citep{pairolero2025aipd}. The analyzed extracted CSV contains 15,437,542 publication-document rows (7,139,652 in 2014--2023); the release records its exact filename, 2,975,669,657-byte size, and SHA-256 \texttt{15954117\ldots734EAE}. The original ZIP hash and exact retrieval timestamp were not retained and are explicitly null, rather than inferred from local timestamps.

For each publication month, the primary panel sums continuous scores for machine learning, evolutionary computation, natural-language processing, speech, vision, planning/control, knowledge processing, and AI hardware across grants and pre-grant publications. This unit is classified publication-document activity, not unique inventions or patent families. Sensitivities separate grants from pre-grants and divide score sums by monthly document count. A family-normalized claim is not made because a validated family identifier is absent from the released aggregate.

Every primary query receives the common eight-component block, which tests generic national AI-publication activity. A released hypothesis-based sensitivity maps each query to named components before evaluation (e.g., \emph{facial recognition}$\rightarrow$vision; \emph{self-driving car}$\rightarrow$vision, planning, hardware). The mapping and rationale are inspectable and were not chosen from evaluation effects.

\subsection{Versioned, tamper-evident artifact and access}

The public repository is \url{https://github.com/pozapas/SignalAudit-QD}; the reviewed snapshot is the tagged \texttt{v1.0.1} release. A fresh run writes project outputs only to a new \texttt{reruns/<run-id>/} directory, with Python/pytest caches disabled. It tests the code, reruns all analyses and the adapter validator, generates five figures and three tables, compiles the paper, and writes run/checksum ledgers. Release verification is performed on the pristine downloaded archive before setup; it rejects changed, missing, and undeclared payloads while ignoring VCS and runtime caches. Separate promotion prevents reruns from overwriting reviewed files.

The package releases each origin prediction and paired query--month loss, not only aggregate metrics. Users can replace the learner, external block, or uncertainty method while retaining the target cells. Google Trends files remain under Google terms and attribution; AIPD derivatives carry USPTO acknowledgement; neither is relicensed as MIT. The resource contains aggregate indices and document classifications, not user-linked queries.

\subsection{Record contract and extension surface}

The archive is deliberately relational rather than score-only. A prediction row is uniquely keyed by target vintage, horizon, target month, query, and model. Alongside actual and clipped/unclipped forecasts, it records origin month, selected penalty, response mode, patent width and mapping, training endpoints/count, event-time eligibility, and historical source-vintage availability. A paired-loss row adds ordered signal/removal identifiers and both cell losses. These keys support metric substitution or rejection of unmatched comparisons without rerunning a learner.

\begin{table*}[t]
    \caption{Released record contract. Keys are stable across reruns; manifests bind their concrete bytes to a version.}
    \label{tab:record-contract}
    \centering
    \small
    \setlength{\tabcolsep}{4pt}
    \begin{tabular}{@{}p{0.18\textwidth}p{0.23\textwidth}p{0.27\textwidth}p{0.24\textwidth}@{}}
        \toprule
        Layer & Row key / grain & Audit-bearing fields & Intended extension \\
        \midrule
        Frozen targets & extraction, month, exact query string & request parameters, collection time, zeros, source hash & alternative target vintages or panels \\
        Patent inputs & definition, publication month, component & document type, aggregation unit, source revision/hash & new public signals or semantic mappings \\
        Origin predictions & vintage, horizon, target, query, model & origin, train bounds/count, tuning, PCA, timing/vintage flags & learners fitted on the same cells \\
        Paired losses & prediction key plus ordered model pair & both losses and signed difference & metrics and matched-removal audits \\
        Summaries and ledgers & comparison/configuration or file path & intervals, seeds, hashes, bytes, test/compile status & uncertainty methods and integrity checking \\
        \bottomrule
    \end{tabular}
\end{table*}

New systems emit the documented prediction columns on frozen origin cells and name their signal/removal models to the released validator CLI. Before constructing cell/month losses and intervals, it rejects duplicate or missing keys, unequal targets or training bounds, inconsistent timing/vintage flags, nonfinite forecasts, and incorrect recorded losses. New external signals also declare the latest source month consumed. The contract distinguishes the forecasting system, availability policy, and comparison estimand.

\section{Matched Evaluation Protocol}
\label{sec:protocol}

\subsection{Task, chronology, and primary pair}

Let $y_{q,t}\in[0,100]$ be the within-query interest index and $\mathbf p_t\in\mathbb R_{\geq0}^{8}$ the monthly patent-score sums. A direct $h$-month pseudo-forecast is issued at event-time origin $o=t-h$. The fixed chronology is 2014--2017 warm-up, 2018--2019 development, and 2020--2023 evaluation: 48 one-month targets and 46 three-month targets. The three-month task begins in March 2020 so its first origin is December 2019, after development. These splits organize event-time rows inside later retrospective source vintages; they do not recreate the information downloaded or classified at each historical origin.

Both primary ridge models include $y_{q,o}$ and query lags 1, 2, 5, and 11 plus target-month sine/cosine terms. The signal model additionally uses $\log(1+\mathbf p)$ at origin lags 0, 2, 5, and 11, standardized on training rows and projected by training-only PCA \citep{jolliffe2016pca}. With $\mathbf z^P_o$ absent from removal, the level model is
\begin{equation}
\widehat y_{q,t}=\beta_{0,q}+\boldsymbol\beta_q^\top\mathbf y_{q,o}
+\boldsymbol\delta_q^\top\mathbf z^{P}_{o}+\boldsymbol\eta_q^\top\mathbf c_t.
\label{eq:model}
\end{equation}
Within each expanding-origin pair, the target rows, response, query and calendar design, penalty, clipping, and training-fitted transforms are identical. Forecasts are clipped to $[0,100]$.

With analysis months indexed from zero and maximum origin lookback $L=11$, the primary training targets at origin $o$ are $\mathcal R(o,h)=\{r:h+L\leq r\leq o\}$. Thus the first training targets are January 2015 for $h=1$ and March 2015 for $h=3$, and the last is always the forecast-origin month. Each query is fitted separately; standardization and patent PCA are refitted on $\mathcal R(o,h)$ at every origin.

The ridge grid is $\{10^{-3},10^{-2},\ldots,10^6\}$ with decade steps. Query-only development macro MAE selects $\alpha=100$ for both horizons, bracketed inside the grid; the primary signal pair freezes that value. A separately tuned patent sensitivity also selects 100. The two-PC bottleneck is inherited from the predecessor coupled model and is audited as a fixed legacy specification, not chosen for current evaluation performance or claimed optimal. We additionally evaluate 2, 4, 8, 16, and all 32 standardized patent-lag variables; development selection at the fixed query penalty chooses 8 PCs for $h=1$ and 16 for $h=3$. The repaired audit protocol was designated after predecessor results were available and was not externally preregistered; diagnostics are not labeled confirmatory tests.

\subsection{Host, mapping, timing, and construct diagnostics}

Persistence is a strong reference, so a matched host sensitivity predicts the change $y_{q,t}-y_{q,o}$ and adds the forecast residual back to $y_{q,o}$. Its query-only penalty is selected on development and then frozen for the patent/removal pair. We also report separately optimized query and patent versions, clearly distinguished from the fixed-penalty signal estimand. Untuned references are last observation and the value 12 months before the target.

Timing controls use row-specific publication months. Added delays $d\in\{1,3,6,12\}$ replace the event-time lags by $d+\{0,2,5,11\}$; each delay receives a removal model with the same later training start. Across delays, that start changes, so the curve is explicitly a joint delay-plus-history sensitivity, not a pure latency effect. The 24-month stress test uses the non-wrapping lag set $24+\{0,2,5,11\}$ and drops unavailable leading rows. A future oracle aligns patent features to the target and is event-time ineligible. No variant uses circular rolling or proves historical classification availability.

Feature-construct checks compare the common block with the released query-component map, and compare score sums for all documents, grants only, pre-grants only, and mean score per monthly document. For the positive control, $r_o=\log(1+\sum_jp_{j,o})-\log(1+\sum_jp_{j,o-2})$ and $s_t=(r_{t-1}-\bar r_D)/\sigma_D$, where $D$ ends in December 2019 and $\sigma_D$ uses population variance. We add $\gamma s_t$ to every query's monthly panel, clip to $[0,100]$, freeze the query-selected $\alpha$, and refit both models at every origin for $\gamma\in\{0,.25,.5,1,2\}$. This synthetic assay asks which event-time-eligible effects the pipeline recovers; it is not evidence about real patent influence.

\subsection{Estimands, uncertainty, and practical scale}

For $Q=20$ and $T_h$ targets, the designated primary contrast is
\begin{equation}
\Delta_h=\frac{1}{T_h}\sum_t\frac{1}{Q}\sum_q
\left(|e^{P}_{q,t}|-|e^{0}_{q,t}|\right),
\end{equation}
where negative favors patents. Macro averaging gives every query equal weight, but a high-error query can still contribute a larger numerical loss. We therefore also archive query-wise RMSE and 12-month seasonal MASE, low-base exclusion, and leave-one-query-out ranges.

Each target month contributes its complete paired 20-query vector. The primary conditional 95\% interval uses 5,000 deterministic circular moving-block replicates with three-month blocks \citep{kunsch1989bootstrap,politis1992circular}; block lengths 1, 6, and 12 are reported as sensitivities. It quantifies stability over this fixed 48-month loss-difference sequence, conditional on the chosen queries, target extraction, patent reconstruction, model, and period. It does not cover query selection, Google sampling or renormalization, patent-classification revision, or future periods. The procedure assumes a roughly stationary, locally dependent sequence and introduces end-to-start adjacency. A non-circular exact sign randomization over the four calendar-year clusters is therefore added, with explicitly coarse resolution. RMSE, MASE, and low-base intervals use the same three-month paired block design. Intervals crossing zero are not equivalence tests. For interpretation only, $\pm1\%$ of persistence MAE defines a descriptive practical band; it was not preregistered as a decision margin.

\subsection{Offline interface: within-query high-interest alerts}

For each query separately, the 75th percentile of its own observations through 2019 is frozen as a high-interest threshold. At each evaluation pseudo-origin, the interface flags whether the next-month forecast exceeds that threshold; a system could use the flag when deciding whether to refresh temporal completion/ranking features. The realized label uses the next value and the same threshold. We macro-average balanced accuracy over the 18 queries containing both classes; F1 covers all 20 with a zero score when its denominator is zero. This single operating point never ranks normalized levels across queries and does not measure refresh cost, user utility, latency, or relevance.

\section{Benchmark Results and Diagnostics}
\label{sec:results}

\subsection{Forecast levels and designated contrast}

Persistence obtains one-month macro MAE 2.953 (Table~\ref{tab:models}). The level query ridge is weaker at 3.150, and adding the designated two-PC patent block yields 3.151. Their paired difference is $+0.0012$ with conditional 95\% interval $[-0.0363,+0.0387]$ (Figure~\ref{fig:signal-audit}). Block lengths 1, 6, and 12 give $[-0.0413,+0.0404]$, $[-0.0370,+0.0350]$, and $[-0.0319,+0.0297]$. Every interval crosses zero; the canonical interval also crosses the descriptive $\pm0.0295$ practical band, supporting neither aggregate improvement nor equivalence.

\begin{table}[t]
    \caption{Fixed 2020--2023 point performance (equal-query macro averages; lower is better). ``Patent $\alpha$'' identifies the removal refit at the penalty selected by the patent-augmented anchored model.}
    \label{tab:models}
    \centering
    \scriptsize
    \resizebox{\columnwidth}{!}{\begin{tabular}{lrrrr}
\toprule
& \multicolumn{2}{c}{1 month} & \multicolumn{2}{c}{3 months} \\
\cmidrule(lr){2-3}\cmidrule(lr){4-5}
Model & MAE & RMSE & MAE & RMSE \\
\midrule
Last observation & 2.953 & 4.557 & 4.402 & 6.612 \\
Seasonal last & 5.347 & 7.899 & 5.475 & 8.030 \\
Query-only ridge & 3.150 & 4.743 & 4.447 & 6.953 \\
Ridge + patents & 3.151 & 4.754 & 4.408 & 6.925 \\
Persistence-anchored ridge & 2.977 & 4.568 & 4.492 & 7.065 \\
Anchored ridge + patents & 2.978 & 4.568 & 4.486 & 7.049 \\
Anchored removal (patent $\alpha$) & 2.943 & 4.535 & 4.492 & 7.065 \\
Anchored + patents (tuned) & 2.948 & 4.541 & 4.486 & 7.049 \\
\bottomrule
\end{tabular}
}
\end{table}

The persistence-anchored host is much closer to the reference: 2.977 query-only and 2.978 with patents at the frozen query-selected penalty. Its matched difference is $+0.00051$ $[+0.00024,+0.00079]$: precisely estimated because the large penalty makes the forecasts nearly identical, but far inside the descriptive practical band. Allowing the patent model to select $\alpha=1000$ yields MAE 2.948, yet its no-patent refit at the same penalty is better at 2.943. Thus the apparently competitive optimized pipeline is not evidence that the patent block caused the gain.

To test reuse outside the ridge implementation, the release includes a nonlinear $k$-nearest-neighbor adapter. Query-only development data select $k=3$, which is frozen for the matched signal/removal pair. Patents worsen its evaluation macro MAE by $+0.2545$ points, with conditional interval $[+0.0535,+0.4663]$. This host-specific harm is not a universal learner claim; it demonstrates that the documented prediction-key and paired-loss contract is independently executable and can reveal a result unlike the ridge estimate.

\begin{figure*}[!t]
    \centering
    \includegraphics[width=0.98\textwidth]{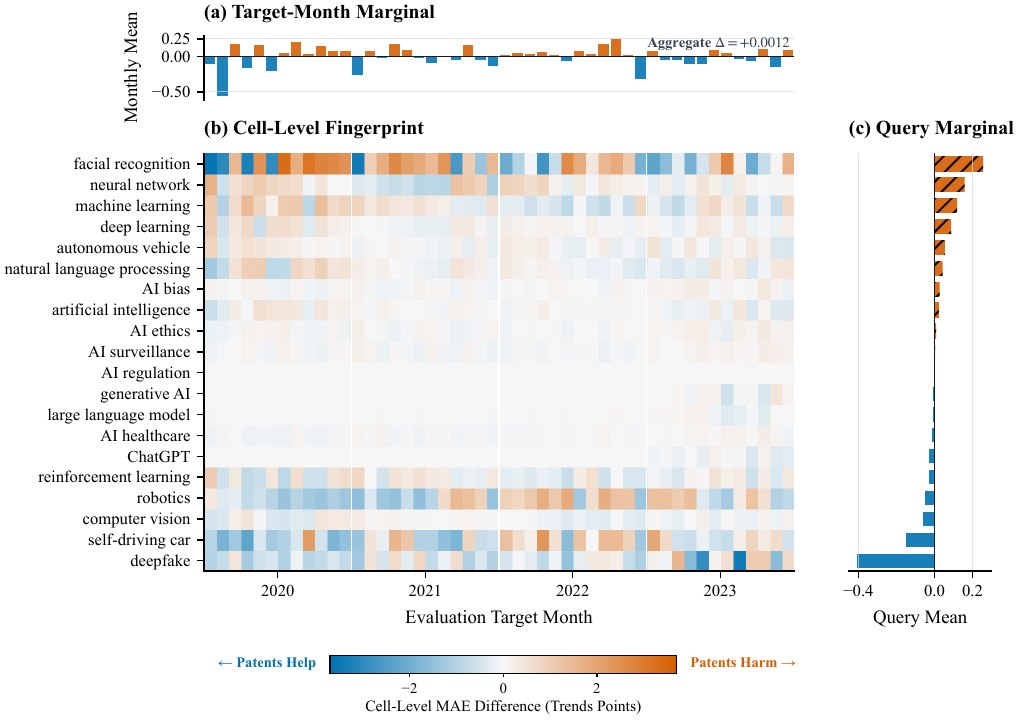}
    \caption{Cell-level fingerprint of the designated one-month contrast across all 960 query--month targets. Each cell is patent-minus-removal absolute error; blue favors patents and orange favors removal. Monthly and query marginals reveal cancellation around the $+0.0012$ aggregate. Cells and sorted marginals are descriptive, not individual tests.}
    \Description{A heatmap with 20 query rows and 48 monthly columns, a monthly marginal bar chart above, and sorted query marginal bars at right. Blue and orange gains and harms coexist around a near-zero average.}
    \label{fig:cell-fingerprint}
\end{figure*}

The near-zero average is not forecast identity. Patents reduce absolute error in 421 cells, increase it in 431, and tie in 108. Monthly means are negative in 20 targets and positive in 28; query means split 10 to 10 (Figure~\ref{fig:cell-fingerprint}). Leave-one-query-out estimates range from $-0.0122$ to $+0.0228$, so no single query creates the full-panel point estimate.

\subsection{Feature, timing, and construct sensitivity}

Development selects 8 PCs at one month. That model's matched effect is $-0.0333$ $[-0.2012,+0.1470]$; retaining all 32 patent-lag features gives $+0.0671$ $[-0.1500,+0.2809]$. The wider intervals show that the small panel cannot sharply resolve high-dimensional patent effects. The selected-8-PC alias and fixed-8-PC row have identical loss vectors and, under common bootstrap draws, identical intervals. The transparent query-component map returns $-0.00045$ $[-0.0377,+0.0369]$, essentially the designated conclusion.

The joint delay-plus-history curve is unresolved: added delays of 1, 3, 6, and 12 months give point effects $-0.0126$, $-0.0247$, $-0.0156$, and $-0.0307$, and every interval crosses zero. Because training starts move with delay, this is not a pure latency contrast and does not rule out latency as a class. The event-time-ineligible future oracle gives $+0.0045$ $[-0.0261,+0.0333]$. Replacing circular roll with a non-wrapping row-wise lag and matching its shorter history changes the 24-month stress result to $-0.0415$ $[-0.0995,+0.0193]$; it is interpreted only as a lag sensitivity.

\begin{table}[t]
    \caption{Canonical one-month audit; every row is compared with its designated row- and penalty-matched removal. Negative favors patents.}
    \label{tab:audits}
    \centering
    \scriptsize
    \resizebox{\columnwidth}{!}{\begin{tabular}{lrr}
\toprule
Patent input & $\Delta$ MAE & Conditional 95\% interval \\
\midrule
Observed patents & +0.0012 & [-0.0363, +0.0387] \\
Dev-selected patent PCs & -0.0333 & [-0.2012, +0.1470] \\
Full patent block & +0.0671 & [-0.1500, +0.2809] \\
Query-mapped patent block & -0.0005 & [-0.0377, +0.0369] \\
Persistence-anchored host & +0.0005 & [+0.0002, +0.0008] \\
Anchored host, tuned patent model & +0.005 & [+0.002, +0.008] \\
Future-patent oracle & +0.0045 & [-0.0261, +0.0333] \\
24-month lag stress test & -0.0415 & [-0.0995, +0.0193] \\
\bottomrule
\end{tabular}}
\end{table}

Changing the patent unit does not produce a clear benefit (Table~\ref{tab:sensitivities}). Relative to $+0.0012$ for all-document score sums, grants-only gives $+0.0126$, pre-grants-only $-0.0092$, and mean score per document $-0.0302$; all intervals cross zero. Excluding the eight queries with at least 50\% pre-evaluation zeros moves $\Delta$MAE to $+0.0384$ over 12 terms, with conditional interval $[-0.0252,+0.0987]$. The exclusion remains unresolved and reinforces the sampling-frame limitation.

\begin{table}[t]
    \caption{Metric, low-base, and patent-definition sensitivities for the canonical one-month contrast. Intervals use three-month moving blocks over paired monthly losses. MASE excludes three zero-scale queries ($Q=17$); other rows use all 20 unless noted.}
    \label{tab:sensitivities}
    \centering
    \scriptsize
    \resizebox{\columnwidth}{!}{\begin{tabular}{llrr}
\toprule
Sensitivity / scope & Metric & $\Delta$ & Conditional 95\% interval \\
\midrule
Exclude low-base queries ($Q=12$) & MAE & +0.0384 & [-0.0252, +0.0987] \\
$Q=20$ & Macro RMSE & +0.0106 & [-0.0386, +0.0468] \\
$Q=17$ & Macro MASE & -0.0083 & [-0.0309, +0.0147] \\
Patent grants only & MAE & +0.0126 & [-0.027, +0.050] \\
Pre-grants only & MAE & -0.0092 & [-0.049, +0.030] \\
Mean score per document & MAE & -0.0302 & [-0.087, +0.024] \\
\bottomrule
\end{tabular}}
\end{table}

Macro RMSE ($Q=20$) changes by $+0.0106$ $[-0.0386,+0.0468]$ and seasonal MASE ($Q=17$) by $-0.0083$ $[-0.0309,+0.0147]$. These intervals and the low-base interval use the same moving-block design as the primary analysis. The non-circular calendar-year sign randomization gives two-sided $p=1.0$; with only four year clusters this is a coarse robustness check, not high-powered inference.

\subsection{Positive control and assay calibration}

At $\gamma=0$, the positive control reproduces the $+0.0012$ primary effect. Planted effects of 0.25, 0.5, and 1.0 points per pre-evaluation standard deviation remain unresolved. At 2.0 points, the matched effect becomes $-0.0728$ $[-0.1307,-0.0189]$ (the bottom calibration row of Figure~\ref{fig:signal-audit}). The pipeline can therefore recover a sufficiently strong event-time-eligible synthetic signal, but the assay is not sensitive enough to validate small effects in this panel. This calibration replaces the unsupported inference that an arbitrarily destructive transform proves sensitivity.

\subsection{Two-pull and three-month boundaries}

\begin{figure}[!t]
    \centering
    \includegraphics[width=\columnwidth]{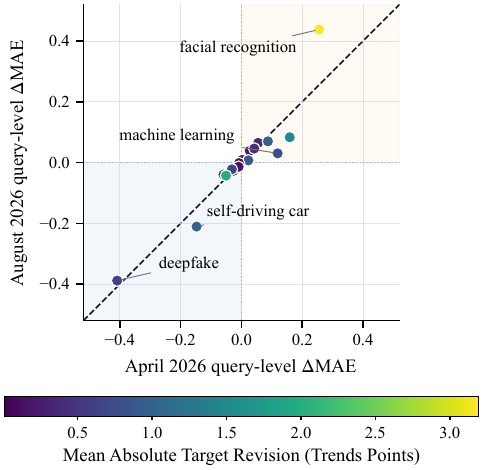}
    \caption{Query-level patent-minus-removal $\Delta$MAE across the two archived target extractions. Color is each query's mean absolute cross-pull target difference. The agreement is descriptive robustness across two pulls, not an estimate of historical revision.}
    \Description{A scatter plot compares April and August query-level effects. Most points lie near the diagonal and in the same-sign quadrants; color shows how much each target series differs between pulls.}
    \label{fig:query-vintage}
\end{figure}

Across 2,400 common 2014--2023 query--month cells, 59.6\% are identical between the April and August pulls; their mean absolute difference is 0.686 points. These differences conflate sampling, normalization, and possible revision. Nevertheless, the designated one-month effect repeats: $+0.0014$ $[-0.0388,+0.0428]$ in August versus $+0.0012$ $[-0.0363,+0.0387]$ in April. Query-level effects have Pearson $r=0.936$, Spearman $\rho=0.977$, and matching signs for 19 of 20 terms (Figure~\ref{fig:query-vintage}).

At three months, level ridge MAE moves from 4.447 without patents to 4.408 with them: $\Delta=-0.0398$, with conditional 95\% interval $[-0.1071,\allowbreak +0.0304]$. In August, $\Delta=-0.0385$, with interval $[-0.1048,\allowbreak +0.0314]$. This remains an unresolved boundary result, not a selected positive horizon.

\subsection{Within-query high-interest alert}

At the frozen 75th-percentile operating point, persistence gives macro balanced accuracy 0.766 ($Q=18$) and F1 0.764 ($Q=20$). The query ridge scores 0.689 and 0.702, respectively; the patent ridge scores 0.685 and 0.702. Two queries lack both classes and are excluded from balanced accuracy. The patent ridge does not improve this particular offline alert, and persistence is strongest. The value is the reusable interface, thresholds, and predictions, not an online-system claim.
\section{Scope, Limitations, and Recommended Use}
\label{sec:scope}

SignalAudit-QD supports a narrow retrospective pseudo-origin statement. The code prevents post-origin event rows from entering event-time-eligible fits, but both Trends targets were downloaded in 2026 and AIPD classifications were reconstructed after many modeled origins. Event time, first public source-release time, classification-version time, and retrieval time are distinct; the archive records the available distinctions and explicitly marks historical source-vintage availability as false. A historically deployable simulation would require prospectively archived query and patent-feature vintages.

The panel is a fixed purposive U.S. AI case study. Its exact strings were fixed before this benchmark's outcomes, but the predecessor project did not preserve a reconstructible candidate universe or selection flow. The query catalog, eight low-base terms, exclusion result, and leave-one-out range bound what can be learned; they do not create representativeness. Independent Trends requests also prohibit cross-query absolute-volume interpretation.

Patent scores are model outputs over publication documents. Score sums mix document volume and composition; grants/pre-grants and mean-per-document sensitivities narrow but do not eliminate this construct uncertainty. AIPD publication does not measure invention, media exposure, adoption, or causal influence. The common national block and transparent query-component map are both coarse. Query-specific retrieval over patent text and validated family-level aggregation remain stronger future designs.

The primary moving-block interval is conditional on the fixed panel, source reconstructions, 48 targets, and model; it relies on approximate stationarity and weak dependence. It is not uncertainty over a population of AI queries, future periods, or source revisions. Alternative block lengths, a coarse non-circular year-cluster test, moving-block RMSE/MASE and low-base intervals, and the practical reference band are diagnostics, not proof of equivalence. The positive control shows detection at the largest planted effect tested, while smaller injections remain unresolved. Likewise, the high-interest alert is an offline within-query interface at one threshold; no cache cost, user interaction, relevance judgment, or online ranking outcome is observed.

Recommended reuse is to compare new learners or external blocks on the released pseudo-origin cells, rerun uncertainty from paired losses, or preregister a broader query sampling frame before collection. The documented adapter requires the stable prediction key, forecast, training bounds, event-time eligibility, and historical-vintage flag; an included nonlinear nearest-neighbor example exercises this interface independently of ridge. The resource should not support individual-level inference, absolute demand allocation, causal claims, or universal conclusions about patents and attention.

\paragraph{Generative-AI disclosure.}
Large language models were used for grammar checking and for rephrasing selected sentences. The authors reviewed and approved all resulting text and remain responsible for the manuscript.

\paragraph{Reproducibility statement.}

The public \texttt{v1.0.1} artifact at
\url{https://github.com/pozapas/SignalAudit-QD/releases/tag/v1.0.1}
contains frozen inputs, hashes, paired evidence, selected configurations,
direct Python pins, realized Python/TeX tool logs, executable tests,
figure/table generators, per-file terms, and run/release manifests.
The one-command workflow writes a new non-overwriting, manifest-bound run
and rebuilds every manuscript asset without changing the reviewed snapshot.

\section{Conclusion}
\label{sec:conclusion}

SignalAudit-QD turns an external-signal claim into a versioned, row-matched object that can be inspected beyond one score. In this fixed retrospective panel, the designated AI-patent block adds no clear aggregate one-month value beyond query history. Feature width, semantic mapping, patent-unit variants, a second extraction, and three months remain unresolved; the anchored and nonlinear hosts instead show model-specific worsening, only a few thousandths of a point for the anchored fits. Correcting the circular lag also removes the earlier apparent harmful control, demonstrating why executable timing tests matter. The positive-control curve supplies the complementary result: the assay detects a sufficiently strong event-time-indexed planted signal but is not powered to certify small effects. The contribution is the public audit resource, paired evidence, and honest diagnostic boundary---not a new forecaster, a historical deployment simulation, or a claim that patents never matter.

\bibliographystyle{plainnat}
\bibliography{references}

\end{document}